\documentclass[12pt]{article}
\usepackage{upgreek}
\usepackage{mathrsfs}
\usepackage{amsmath,amssymb,color}
\usepackage{amsfonts}
\usepackage{url}
\usepackage{xcolor}
\usepackage[colorlinks,linkcolor=black,anchorcolor=blue,citecolor=blue,urlcolor=blue]{hyperref}
\usepackage{graphicx}
\usepackage{wrapfig}
\usepackage{cases}
\usepackage{appendix}
\usepackage[T1]{fontenc}
\usepackage{authblk}
\usepackage{txfonts}
\usepackage[numbers,sort&compress]{natbib}
\makeatletter
\@addtoreset{equation}{section}
\makeatother

\newcommand{\er}[1]{\eqref{#1}}

\newcommand{\ci}[1]{}

\newcommand{\ke}{\rangle}
\newcommand{\br}{\langle}

\newcommand{\nn}{\nonumber \\}

\newcommand{\ba}{\begin{eqnarray}}
\newcommand{\ea}{\end{eqnarray}}
\newcommand{\be}{\begin{equation}}
\newcommand{\ee}{\end{equation}}
\newcommand{\bal}{\begin{align}}
\newcommand{\eal}{\end{align}}
\newcommand{\bay}[1]{\left(\begin{array}{#1}}
\newcommand{\eay}{\end{array}\right)}
\newcommand{\eg}{\textrm{e.g.} }
\newcommand{\ie}{\textrm{i.e.}, }

\newcommand{\st}[1]{|#1\ke}

\def\rmd{{\rm d}}

\def\xb{{\beta}}

\def\xl{{\lambda}}

\def\xO{{\Omega}}

\def\xs{{\sigma}}

\def\xt{{\theta}}

\def\CC{{\cal C}}

\def\CO{{\cal O}}

\def\CS{{\cal S}}

\title{Growth of a renormalized operator as a probe of chaos}
\author[a,b,c]{Xing Huang \thanks{xingavatar@gmail.com}}
\author[a]{Binchao Zhang \thanks{zhangwusanbinchao@163.com}}
\affil[a]{Institute of Modern Physics, Northwest University, Xi'an 710069, China}
\affil[b]{NSFC-SPTP Peng Huanwu Center for Fundamental Theory, Xi'an 710127, China}
\affil[c]{Shaanxi Key Laboratory for Theoretical Physics Frontiers, Xi'an 710069, China}

\date{}
\begin{document}

\maketitle
\begin{abstract}
We propose that the size of an operator evolved under holographic renormalization group flow shall grow linearly with the scale and interpret this behavior as a manifestation of the saturation of the chaos bound. To test this conjecture, we study the operator growth in two different toy models. The first is a MERA-like tensor network built from a random unitary circuit with the operator size defined using the integrated out-of-time-ordered correlator (OTOC). The second model is an error-correcting code of perfect tensors, and the operator size is computed using the number of single-site physical operators that realize the logical operator. In both cases, we observe linear growth.

\end{abstract}

\newpage
\noindent\rule[0.1\baselineskip]{\textwidth}{0.8pt}
\tableofcontents
\noindent\rule[0.1\baselineskip]{\textwidth}{0.8pt}
\linespread{1.5}\selectfont

\section{Introduction}
In the seminal work by Maldacena, Shenker and Stanford \cite{Maldacena:2015waa}, it was argued that in a large $N$ chaotic theory,
the OTOC is bounded by the exponential growth with the Lyapunov exponent $\xl_L \le 2 \pi/\xb$. The bound is saturated when the theory admits a gravity dual.
This conjecture has been verified in numerous cases. The bound on chaotic
behavior also leaves its mark on various other measures like the spectral
form factor \cite{Cotler:2017jue}, growth of operator \cite{Roberts:2018mnp,Qi:2018bje}, pole skipping \cite{Blake:2017ris,Blake:2021wqj} etc.
 
It has long been speculated that gravity shall emerge from the entanglement structure, but the precise procedure remains unclear. One way to build up the geometry is to consider the entanglement renormalization, which is essentially the radial evolution of the wave function. In some discrete many-body systems the wave function can be expressed in terms of a tensor network, which consists of two types of local operations (disentangler, isometry) that modify the entanglement structure (see \eg \cite{vidal2009entanglement} for a review). Originally, this approach with the name of multi-scale entanglement renormalization ansatz (MERA) was designed for solving the ground state wave function. It was later realized \cite{Swingle:2009bg, Swingle:2012wq} that the MERA tensor network bears a lot of similarities with the AdS space and hence could be understood as a concrete realization of emergent spacetime. The discrete MERA was later generalized \cite{Miyaji:2015yva, Miyaji:2016mxg} to the continuous situation of AdS/CFT (cMERA). Each layer of the tensor network becomes a codimension-two surface in the bulk and corresponds to a state in the CFT (surface/state correspondence). States at different scales are related by unitary radial evolution operators $W$, which in turn define the renormalized operators
$W^\dagger \CO W$ (as if in the Heisenberg picture).  

It is natural to expect the same chaotic behavior for the radial evolution as if the latter is viewed as a dynamic process. In this sense, the fact that the chaos bound is saturated in the gravity system makes it an ideal probe for spacetime geometry. For example, the renormalized operator grows along the radial direction and shall have the same chaotic behavior as in the evolution along the time direction. The chaos bound can then tell whether the wave function corresponds to some bulk geometry. In other words, chaos indicates how information is scrambled under the RG flow and hence greatly influences the entanglement structure.

The state or wave function in the continuous MERA network follows from the Euclidean path integral
defined on the codimension one surface \cite{Caputa:2017urj, Caputa:2017yrh}, which is the same path integral of the boundary CFT (up to a conformal factor). Due to conformal symmetry, it is natural to expect the same dynamics and hence the same chaotic behavior.

However, establishing the measures of chaos along the renormalization group flow is not a simple task. It is not that we can remake all the known results by simply replacing the time coordinate with its radial counterpart. One of the difficulties is that we don't have a clear definition of temperature for the states generated by entanglement renormalization. The UV state we consider is the vacuum and has zero temperature, at which the chaos bound becomes trivial. Moreover, the radial ``time'' parameterizing the entanglement
renormalization is ambiguous and obviously different choices affect the growth. In the presence of a holographic dual, the radial proper distance or its function is a good candidate. In general we may use the energy scale as in ordinary RG flow.

We get some hints from the recently proposed size-momentum correspondence \cite{Susskind:2019ddc, Susskind:2020gnl}, which can be applied to the vacuum case. It was found that the size and complexity of an operator shall grow linearly in time or the circuit time. The operator in this scenario is evolved along the time direction with the ordinary Hamiltonian. It is not a renormalized operator even though it corresponds to a point particle in the bulk. Despite the difference in the physical meaning, we believe the same relation holds up in our scenario. In the rest of the paper, we will focus on the growth of the size of an operator in the radial direction and then establish the chaos bound in a gravity theory. We propose that the size and also the corresponding complexity of an operator cannot grow faster than the characteristic energy scale or rather the radial coordinate of the operator. Supporting evidence is found from some toy models of tensor networks that are decent approximations of discrete AdS space. We observe linear growth in two types of tensor networks, one built from random unitary circuit \cite{Nahum:2016muy, Keselman:2020fmo} and the other being the famous error-correcting code made of perfect tensors \cite{Pastawski:2015qua}. 
 
The rest of the paper goes as follows. In section~\ref{sec:sizemomentum},
we review the idea of size-momentum correspondence and explain its implications
in our scenario. In section~\ref{sec:randomcircuit}, we use a simple toy model of a random unitary circuit to demonstrate how our idea works. More precisely we use the circuit to turn an unentangled IR state into a UV state that emulates the ground state of a holographic CFT in the sense that the Ryu-Takayanagi (RT) formula is reproduced. We then show that the integrated OTOC or equivalently operator size grows linearly with the circuit time or rather the radial coordinate. In section~\ref{sec:errorcorrection}, we investigate the operator growth in
the pentagon error-correcting code based on perfect tensor and show that the size of a renormalized operator does grow linearly. Finally in section~\ref{sec:discussion}, we conclude and discuss about future directions.

\section{Size-momentum correspondence}
\label{sec:sizemomentum}
It was conjectured that \cite{Susskind:2019ddc, Susskind:2020gnl} the size
of an operator corresponds to the radial momentum of a particle in the bulk
and hence the dynamics of the latter is governed by the evolution of the
size or equivalently the complexity, whose growth rate with respect to the
circuit time in turn gives the size. The readers should be aware that there are other definitions of size. In SYK model it measures the number of fermions
needed to composite the operator \cite{Roberts:2018mnp, Qi:2018bje}. We believe
all these definitions as in the case of complexity shall agree for practical
purpose but their precise connection remains unclear. We will review the basic results in \cite{Susskind:2020gnl} and offer our new observations from a different perspective. In this case, the size $s$ is defined in the following
way 
\be\label{sizedef} s(\tau) = \frac {d \CC}{d \tau}\,,\ee
where $\CC$ is the complexity and $\tau$ is the circuit time. When a
gravity dual is assumed, $\tau$ is simply identified as the radial proper
distance $\rho$ in the AdS$_{d+1}$ space
\be
\rmd s^2 = -\cosh^2 \rho \rmd t^2 + \rmd \rho^2 + \sinh^2 \rho \rmd \xO_{d-1}^2,\quad
\rho \in [0,+\infty), t \in (-\infty,+\infty)\,.
\ee
It was then argued that $s$ shall go as
\be s(\rho) \sim e^{\rho}\,.\label{sizegrowth}\ee
The holographic dual of the operator is an infalling particle whose geodesic
gives the radial momentum
\be
2\pi P \equiv \frac {d \CC}{d \rho} = s(\rho) \sim t\,. \label{sizemomentum}
\ee
Now we will make a bold conjecture that such a relation \er{sizegrowth} is understood as the saturation of the chaos bound saturated in gravity system. In this sense, the time evolution of size is a consequence that the particle cannot move faster than light.

The operator growth was known as a measure of chaos \cite{Roberts:2018mnp, Qi:2018bje} and a bound similar to the case of OTOC was established for
finite temperature. At first glance our proposal appears to be quite different. To see the connection, we first note that the near horizon region of a black hole (BH) can always rewritten in the Rindler
coordinate, in which the BH time $\tilde t$ is related to the time $t$ of a vacuum space as 
\[e^{\frac {2\pi } \xb \tilde t} \sim t\,.\]
In other words, we recover the well-known results of chaos bound as long as the size grows linearly with $t$ in a vacuum state.

The reader shall be aware that an operator in the size-momentum correspondence is still evolved by the Hamiltonian in the time direction. On the other hand, the fact that its holographic dual is a local object makes its very tempting to think that a renormalized operator, which in the tensor network is acting on a bulk site will have the same growth in the radial direction.

For later convenience, we replace the radial proper distance by the characteristic length scale $z$ (in pure AdS, $\rho \sim \log z$), which is simply the radial
coordinate in the Poincare patch 
\be
\label{AdSPmetric}
   \rmd s^2=\frac{\rmd z^2-\rmd t^2+\rmd \vec x^2}{z^2}.
\end{equation}
It also measures the scope of influence of the renormalized operator, which is associated with a causal diamond with a spherical RT surface of radius
$z$. Moreover,
we note that from the definition
of \er{sizedef}, the size and complexity share the same behavior of growth.
As a result, we will henceforth take them as interchangeable and sometimes loosely
refer to their behavior as growth of operator. The bound on chaos then implies the size or complexity of a renormalized operator shall grow at most linearly in $z$.   

\section{Random unitary circuit as a discrete AdS space}
\label{sec:randomcircuit}
\begin{figure}[htbp]
  \centering
\includegraphics[width=10.5cm]{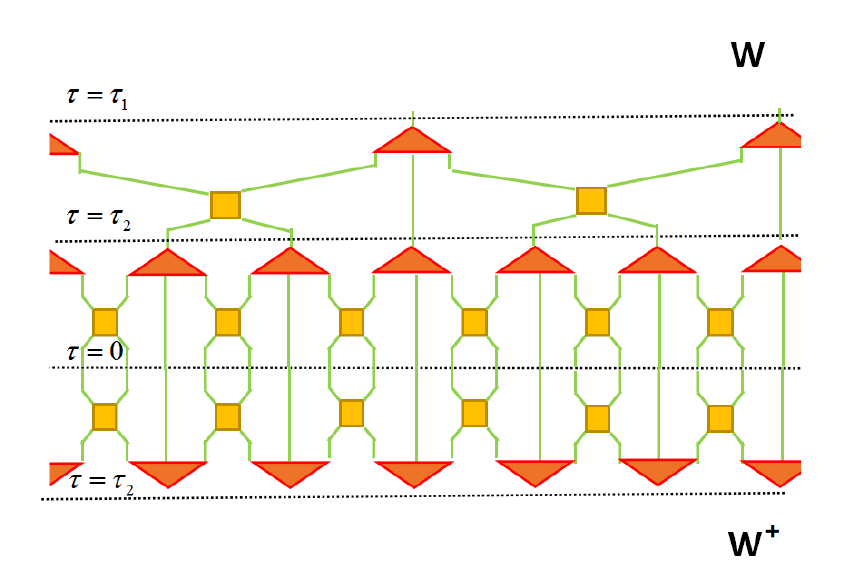}
  \caption{The MERA tensor network is a circuit $W$ made of disentangler (boxes) and isometry (triangles) operations. More precisely $W(\tau_2,\tau_1)$ is given by the product of all the tensors between the lines $\tau=\tau_1$ and $\tau=\tau_2$ and provides a map from low-energy states ($\tau=\tau_1$) to high-energy states ($\tau=\tau_2$). The Euclidean times $\tau_1, \tau_2$ again denote different energy scales and $\tau=0$ corresponds to the UV cutoff. }\label{tensorW}
\end{figure}

Now we would like to study the operator growth in some specific models. Ideally it would be nice to work in the actual AdS/CFT setup but now we have to settle
for some discrete toy models. The goal is to study the growth of a renormalized operator on a tensor network that serves as a good approximation of a spatial slice of the AdS$_3$ space. For starters, let us be clear about the definition of the renormalized operator. In the MERA tensor work (see Fig.~\ref{tensorW}), two types of local unitary operations (gates) are applied to the state. A disentangler removes the local entanglement and an isometry performs coarse graining. Eventually all the spatial entanglements are removed. The collection of all these gates compose a circuit, which provides a map $W(\tau_2, \tau_1)$ between Hilbert spaces at different scales. So a simple IR operator $\CO$ can be evolved radially to UV as $W^\dagger \CO W$ (for simplicity we denote the map from IR to UV as $W$), which becomes in general a complicated non-local operator as in the time-like evolution.

There are various attempts to simulate the discrete AdS space using tensor networks. However, we prefer to keep the size of Hilbert space fixed at each step of the RG flow, in agreement with the surface/state correspondence. Therefore, we first study a toy model built from random unitary circuit and get back to the pentagon code in the next section. 

Recently people \cite{Keselman:2020fmo} have studied the chaotic behavior of the
so-called integrated OTOC, which is essentially the size of an operator evolved in a random unitary circuit consisting of
two types of quantum gates (SWAP, CNOT) with certain probability. A SWAP gate exchanges the the states of the two qubits next to each other. What we are going to do is to replace their CNOT gate by an Eadd gate that works as the inverse of disentangler. The reason for the inverse is that in our case the circuit goes from IR to UV. As we shall see, this minor change gives a completely different circuit that emulates a MERA network or discrete AdS space. For our convenience, we would like to make the circuit time $t$ correspond to the radial $z$ coordinate. It should be noted that the size of an operator in its definition \er{sizedef} depends on the choice of circuit time. But here the size is taken as a physical quantity and is computed in a different way. Consequently our choice of circuit time has no effect on the size. We will compute the integrated OTOC (taken as the size) and show that it grows linearly with the circuit time $t$. To make sure that the circuit constructed in this way does look like AdS space, we also compute the entanglement entropy of a single interval on the boundary and find that it has the right logarithmic behavior as indicated by RT formula.

\subsection{OTOC}
As in \cite{Keselman:2020fmo}, we consider an 1d lattice of $N$
qubits. We start with an IR state with no spatial entanglement, or more precisely a state
with all qubits being $0$ and invert the RG flow towards UV. Physically, the ``time" evolution of the circuit represents a renormalization group flow from IR to UV.
At each time slice, unitary gates are applied with some fixed probability.
As in the MERA circuit, we need two types of gates in the circuit: the disentangler and the isometry. 

Now the claim is that a SWAP gate, which switches
the states of two adjacent qubits,  works almost as good as coarse graining
(isometry) \footnote{Strictly speaking, the argument below is only valid in a state with all entanglements being bi-partite, \ie taking the form of Bell pairs. However, for the study of entanglement renormalization this picture is actually a decent approximation \cite{huang}.} or rather its inverse. In the MERA network, coarse graining brings together qubits that are entangled but far apart and eventually such entanglement is removed by the disentangler. Following \cite{Keselman:2020fmo}, we divide all the ordered pairs into two groups in the forms of $(2m, 2m+1)$ and $(2m-1, 2m)$ ($m$ being positive integers) respectively. The SWAP gates act on two group alternately. A circuit swaps $(2m, 2m+1)$ pairs at odd steps and $(2m-1, 2m)$ pairs at even steps (or the other way around). Such a particular arrangement guarantees that entangled pairs (of neighboring sites) are gradually stretched out. So the inverse operation of isometry is achieved by the SWAP gates and hence we only need to introduce disentanglers. Since the RG flow is inverted (IR $\to$ UV), what we need are entanglers
that add entanglement to the system. More precisely, this type
of gate (we call ``Eadd") is to add Bell pairs to the spatially unentangled state  
\begin{align}
\st{00} & \rightarrow \frac 1 {\sqrt 2}\left(\st{00}+\st{11}\right),\quad \st{11} \rightarrow  \frac 1 {\sqrt 2}\left(\st{00}-\st{11}\right)\nn
\st{01} & \rightarrow  \frac 1 {\sqrt 2}\left(\st{01}+\st{10}\right),\quad \st{10} \rightarrow  \frac 1 {\sqrt 2}\left(\st{01}-\st{10}\right) \nonumber
\end{align}
In the MERA network, the total number of Bell pairs grows exponentially with the
depth $n$ of the circuit. In the continuous case, we have $z \sim 2^n$ and
hence the total entanglement shall grow linearly in $z$. In other words,
we can take the probability of the Eadd gate to also be a constant if the circuit time is taken to be the $z$ coordinate.
The constructed circuits are different in every run due to randomness but eventually we take
an average of the computed quantities (like correlators and entanglement entropy).

To see the chaotic behavior, we consider the usual OTOC. In this case, it is defined by the commutator of two operators $V,U$ at different circuit
time. At time $t=0$, the simple operator $V_j(0)$ is a Pauli matrix acting only on a single site labeled by $j$. Such an operator initially at site $i$ after the unitary time evolution $W(t) \equiv W(\tau_1,\tau_2)$ (or rather under the renormalization
group from IR scale $\tau=\tau_2$ to higher energy scale $\tau=\tau_1$), becomes a very complicated object $U_i(t) =W^\dagger (t) U_i(0) W(t)$ and can have nonvanishing overlap with operators at different sites. Following \cite{Keselman:2020fmo},
we use the integrated OTOC, which is defined in the following way
\[f(t) = \sum_{j} C_{i,j} (t)= -\sum_{j} \br [U_i (t), V_j (0)]^2 \ke\,.\]
In our case, $V_j(0)$ is taken to be an IR operator and $U_i(t)$ is the renormalized
operator at a higher energy scale obtained by reverting the renormalization group flow. As $V,U$ are both Pauli matrices (even though at different energy scales), they satisfy $V^2 = U^2 = 1$ and we have 
\[[V_j(0)U_i(t)-U_i(t)V_j(0)]^2 = V_j(0) U_i(t) V_j(0) U_i(t) + U_i(t)V_j(0)U_i(t)V_j(0) - 2\,.\]
The commutator is evaluated with respect to the unentangled IR state. This quantity can also be understood
as the size of the operator $U_i(t)$. The expectation value is an
average over all different states generated by different random circuits.
\begin{figure}
\centering
\includegraphics[width=0.75\textwidth]{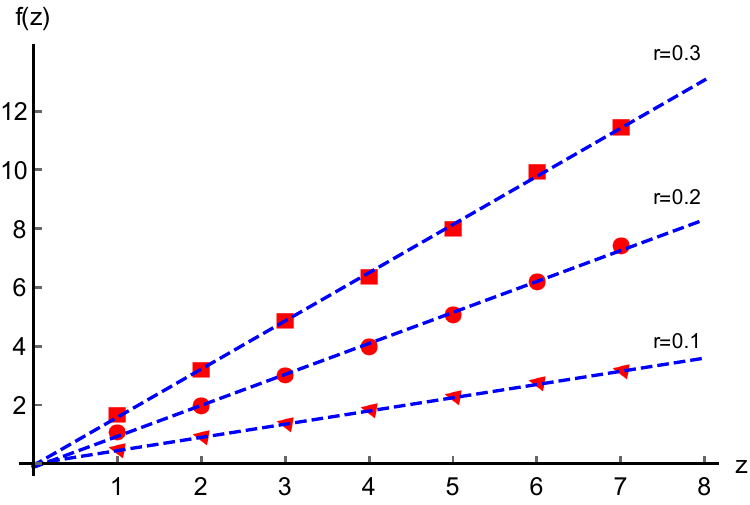}
\caption{Numerical results for the OTOC in a random circuit resembling AdS space. To minimize boundary effect and parity asymmetry, we take the average of the data from $i=8$ and $i=9$. For the cases of $p = 0.9$ and $r=0.1, 0.2, 0.3$ respectively, $f(t)$ is approximately a linear function of radial scale $t$ in AdS space. }
\label{fig:OTOC}
\end{figure}

To match the linear growth in the holographic picture discussed in Sec~\ref{sec:sizemomentum}, we are supposed to go the other way and measure an IR operator
in terms of the simple UV operators. However, the size and complexity of an operator grow exponentially in the proper distance, or linearly in the radial coordinate $z$. In the the current case the OTOC essentially measures the relative complexity between operators at different scales (one being IR) and hence it should be proportional to the difference in $z$ (or equivalently circuit time $t$).
In other words, it should grow linearly in $t$.  
  
For demonstration purpose, we consider a small circuit of 16 qubits ($N=16$).
Two types of gates are applied randomly at every time step in $t$. We set the probability of SWAP gates fixed at $p=0.9$ and consider three different choices for the probability of Eadd ($r = 0.1, 0.2, 0.3$). The coding is done using ITensor package \cite{Fishman:2020gel}. This setup limits our evolution to $7$ time steps ($t=7$). Despite the small number of data points, a linear behavior of OTOC is observed for various $r$ as shown in Fig.~\ref{fig:OTOC}. We note that $f(0) = 4$ (for a trivial circuit with no gates) and hence all the data values in the figure are shifted by $-4$ for better visualization. 
\subsection{Entanglement entropy} 
\begin{figure}
\centering
\includegraphics[width=0.75\textwidth]{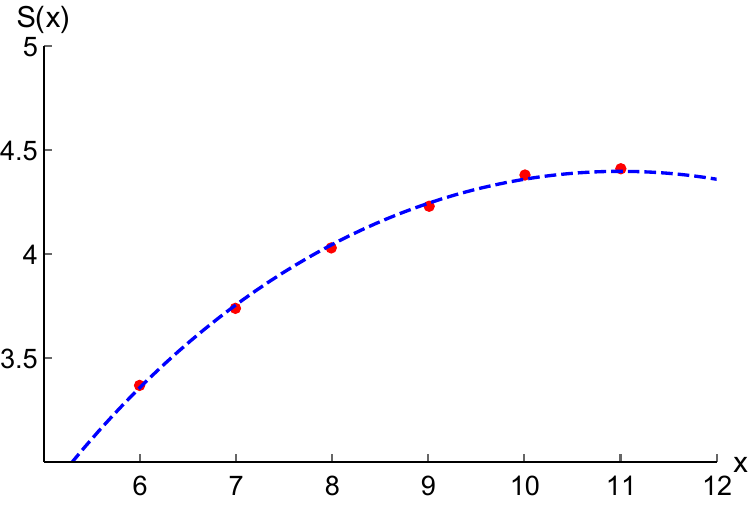}
\caption{Numerical results for the entanglement entropy for the UV state obtained using the random circuit (at $t=8$) with parameters $N=22, p = 0.9,r=0.2$. Entanglement entropy of a subsystem $S(x)$ can
be approximated by a logarithmic function of its size $x$ as required by the RT-formula.}
\label{fig:EE}
\end{figure}

To see that our random unitary circuit does emulate the discrete AdS space,
we check the entanglement entropy of the UV state obtained by running the
circuit to the time $t=8$. To get a slightly better resolution, we extend
the lattice to include 22 qubits. Moreover we impose periodic boundary condition. In this case the grid points are supposed to be labeled by angular coordinate and the Ryu-Takayanagi formula in this case takes the following form
\be
S(\xt_1, \xt_2) = \frac c 3 \log \sin \frac {(\xt_1 - \xt_2)} 2 \,.\label{RTformula}
\ee
With periodic condition, a subsystem with $x$ quibts corresponds to an angle
of 
\[
2 \pi \times \frac x {22} \,.
\]
So we shall try to fit the data using
\[
a \log \sin \left(\frac{\pi  x}{22} \right)+b\,,
\]
where $x$ is the number of qubits in the subsystem and the precise values of $a,b$ are unimportant. We note that the RT formula
is not very accurate for small intervals due to the discrete nature and hence we start with a system of 6 qubits. The entanglement entropy decreases if the size gets past half of the total and hence we stop at $x=11$. As we can see in Fig.~\ref{fig:EE}, the entanglement entropy can be described by the
RT formula with good accuracy.  

\section{Operator growth in a network of perfect tensors}
\label{sec:errorcorrection}
In \cite{Pastawski:2015qua}, a MERA-like tensor network was constructed using
the product of pentagon perfect tensors \footnote{A tensor $T$ is perfect if $T^\dagger T$ gives the identity when half or more of its indices are contracted.}. Such
a network shares many nice features with the AdS space, like RT-formula and error
correction (subregion duality \cite{Jafferis:2015del, Dong:2016eik}).  
We would like to investigate the complexity of a bulk operator in this setup.
More precisely, a bulk operator in the sense of error correction is taken
as a logical operator and it can be mapped by the network to a subsystem
in the physical space on the boundary. In other words, the logical operator can be recovered from a subregion (protected from the erasure of the rest).
The physical operator in general is non-local in its support of the whole subregion and hence has large complexity measured by the single-site operators
(taken to be simple). Since complexity and size grow the same way, we expect the complexity to grow linearly in the radial coordinate.   

First of all, let us consider a single perfect tensor. The logical operator in the 5-qubit (pentagon) code $[[5,1,3]]_2$ is given by the tensor product
of single-site operators ($X,Z$ being Pauli matrices)
\be
\label{logicalop513}
\bar X = X \otimes X \otimes X \otimes X \otimes X,\quad \bar Z = Z \otimes Z \otimes Z \otimes Z \otimes Z\,.
\ee
Here we use the same notation as in \cite{Pastawski:2015qua}. A code denoted
as $[[m,k,d]]_v$ has $m$ physical spins (of degrees of freedom $v$) or sites to encode
$k$ spins, which can be recovered from $d$ spins. The encoding process is realized as a one-to-five map using the 6-qubit
(hexagon) perfect tensor. A logical operator can be realized in various equivalent representations 
\[\bar X \sim \bar X U,\quad \bar Z \sim \bar Z U,\quad U \in \CS.\]
Here $\CS$ is the stabilizer group that consists of the operators (also in the form of a product of single-site Pauli
matrices) keeping the perfect tensor invariant. In the current case, it is generated by $S_i$ with the following explicit form
\[ S_1 = X \otimes Z \otimes Z \otimes X \otimes I\,,\]
\[ S_2 = I \otimes X \otimes Z \otimes Z \otimes X\,,\]
\[ S_3 = X \otimes I \otimes X \otimes Z \otimes Z\,,\]
\[ S_4 = Z \otimes X \otimes I \otimes X \otimes Z\,.\]
As a result, $\bar X$ and
$\bar X U$ act the same way on the perfect tensor. Some representations act only on sites in a certain subregion. For example, 
\[\bar X U = -Z \otimes X \otimes Z \otimes I \otimes I\,,\quad U = S_2 S_3 S_4\,,\] 
which seems to imply its complexity is $3$ as it can be realized by the product of three simple operators. We also check that $[[3,1,2]]_3$ has similar properties.

It is natural to think that for a general stabilizer code $[[2n-1,1,n]]_v$ we have the logical operators all
with complexity $n$. For this code, we can recover the information from $n$
spins and hence there would be a representation of logical operators in the form of tensor product of $n$ Pauli matrices (just like those in
\er{logicalop513}, but in a generalized form for $v$ dimensions). The group of all such products is denoted as $G_n$. The point is that two different
representations (denoted as $\bar X_1, \bar X_2$) of a logical operator are equivalent as
\[\bar X^{-1}_1 \bar X_2\]
keeps all the states in the code space invariant and hence has to be an element of $\CS$.
It is reasonable to assume that $\bar X$ is an element of $G_n$ and then so
is the subregion representation $\bar X U \in G_n$. Such an operator has complexity
$n$ as it is the tensor product of $n$ simple (singe-site) operators.

Moreover, a tensor network composed of perfect tensors is also stabilized
by the Pauli matrices acting on the remaining open (uncontracted) sites. We then expect that the logical operators are also in a similar product form. Therefore the complexity of a bulk operator is bounded by
the size of the one dimensional subregion on the boundary in which it is encoded. The corresponding subregion (\ie entanglement wedge) in the AdS bulk is half of a disk with the radius $z$ (radial coordinate of the bulk operator) and hence we get the same linear growth behavior as before. 

We would like to prove this statement by induction. First of all, we would like to consider the logical operators in a many-to-many code by a single tensor. A perfect tensor generally provides the coding of more than one spins. For example, a perfect tensor with four indices $T_{ijkl}$ (\ie $[[4,0,3]]_3$) can be used for a 3-qutrit code $[[3,1,2]]_3$. At the same time, it can be used to encode an operator acting on two qutrits in the sense that the third qutrit can be included as logical (we denote the original logical qutrit as the 4th). The logical operators in the physical space become 
\be\label{22physical} \bar O_{k'l',kl}=T_{i'j'k'l'}O_{i'j',i''j''}T^*_{i''j''kl}\,,\ee
and is a map between the spaces of two qutrits. 

It is known that the basis in the code space of $[[3,1,2]]_3$
\[\st{\bar i} = T_{i j k l } \st{jkl}\,,\]
takes the following explicit form \cite{Almheiri:2014lwa}
\[\st{\bar 0} =\frac 1 {\sqrt 3}(\st{000}+\st{111}+\st{222})\,,\]
\[\st{\bar 1} =\frac 1 {\sqrt 3}(\st{012}+\st{120}+\st{201})\,,\]
\[\st{\bar 2} =\frac 1 {\sqrt 3}(\st{021}+\st{102}+\st{210})\,.\]
To get a better understanding the 2-to-2 code, one can apply the unitary transformation $U_{12}$ 
\[\st{00}\rightarrow\st{00}\quad\st{11}\rightarrow\st{01}\quad\st{22}\rightarrow\st{02}\]
\[\st{01}\rightarrow\st{12}\quad\st{12}\rightarrow\st{10}\quad\st{20}\rightarrow\st{11}\]
\[\st{02}\rightarrow\st{21}\quad\st{10}\rightarrow\st{22}\quad\st{21}\rightarrow\st{20}\]
on the first two qutrits and turn the basis into the form of
\[\frac 1 {\sqrt 3}\st{i}(\st{00}+\st{11}+\st{22})\,. \]
In other words, the transformation $U_{12}$ turns the 2-to-2 ($34 \to 1 2$) map by $T$ into the identity matrix. Therefore a logical operator $\bar O_{3\otimes 4}$ is realized trivially
\[\bar O_{3\otimes4} = U_{12}^\dagger O_{1\otimes2} U_{12}\,. \]
Such a code (and similarly the one by $T$ without $U_{12}$) even though trivial and not protective against any errors, remains legal and can serve as our starting point. 

As explained earlier, physical operators in the form of \er{22physical} for $\bar O_3$ and $\bar O_4$ both have representations supported in the first two qutrits. Since the (single-site) logical operators on different sites commute with each other, so do their different representations. Even though here we only consider $[[4,0,3]]_3$ as an example, it is not difficult to see that the same conclusion remains valid for other perfect tensors. Now we can continue the induction. When one more perfect tensor $T$ is added to the network $P_n$ to get a new $P_{n+1}$, it has a few physical sites (denoted as $B$) contracted to the physical sites in $P_n$. As a reminder, we reverse the construction order in \cite{Pastawski:2015qua} (now from IR to UV). To maintain the isometry property of the new tensor $P_{n+1}$, we cannot contract more than half of the sites of $T$. Since we only care about the local bulk operator, it is OK to consider one logical site. The logical operator $\bar X$ is equivalent to a representation supported only on $A$ ($A = B \cup C$)
\be
\bar X \sim X_{A}\equiv \bar X U = \xs_B \otimes \dots \otimes \xs_B \otimes \xs_C \otimes \dots \otimes \xs_C \in G_A\,,
\ee
where $G_A$ (like $G_n$ before) is the group for the product of Pauli matrices on sites
in the region $A$. Each local Pauli operator $\xs_B$ (they can be different
for different sites but we use the same notation $\xs_B$ for simplicity) in region $B$ can be taken as a logical operator
$\bar \xs_B$ and is replaced by their own subregion
representation on the new boundary $D$ through the coding by the perfect tensor $T$. So we have
\ba
\bar X \sim X_{B \cup C} & = & \bar \xs_B \otimes \dots \otimes \bar \xs_B \otimes \xs_C \otimes \dots \otimes \xs_C \nn
& = &(\bar
\xs_B \otimes \dots \otimes I) (I \otimes \bar\xs_B \otimes \dots \otimes I) \dots (I \otimes \dots \otimes \xs_C)\nn
& = & (\xs_D \otimes \dots \otimes \xs_D \otimes I\otimes \dots \otimes I) \dots (I \otimes \dots \otimes \xs_C \otimes \dots ) \dots (I \otimes \dots \otimes \xs_C)\,.\quad\quad
\ea 
In the second line, we first decompose the physical operator (in a representation for subregion $A$) as the product of single-site operators. In the last line, single-site operators in $B$ are replaced by products in $D$. The final form of the representation remains an element of $G_{D \cup C}$ and its complexity is given by the total number of sites in the subregion $D \cup C$.

In summary, we show that a logical operator can be constructed as the product of single-site operators whose total number agrees with the distance of the code. We take this number as the complexity of a renormalized operator instead of its size. It should be noted however that given the product structure, the integrated OTOC shall have the same growth.  

\section{Discussions}
\label{sec:discussion}
It is known that the entanglement renormalization understood as radial evolution is governed by the same Hamiltonian as in the time evolution and hence the renormalized operators shall satisfy a similar chaos bound. Motivated by the size-momentum correspondence, we further propose that the size of a renormalized operator shall grow linearly with respect to the radial coordinate in a holographic theory. We study the growth of operator size in two different scenarios. In the first case, we construct a random unitary circuit that produces a tensor network similar to the MERA network, which can be taken as a discrete AdS space. This is supported by the fact that the entanglement entropy of the UV state is given by the RT formula. The size of an operator is measured in terms of the integrated OTOC and is found to grow linearly with the circuit time. In the second case,  we consider the perfect tensor code. A logical operator is understood as a renormalized operator and its size is defined by measuring its physical realization in the sense of error correction in terms of the single-site simple operator. It is found that the physical operator is given by the product of a number of single-site operators. Moreover, such a number is taken as the size and grows linearly with the size of the boundary subregion to recover the logical operator. 

Frankly speaking, our results only provide modest support for the claim that a renormalized operator in a holographic theory grows linearly. It is far from evident that there is a universal chaos bound. In a general theory, entanglement renormalization can always be performed using MERA (or cMERA). Given the locality of both disentangler and isometry operations, it is natural to expect a bound on how quickly the entanglement structure is built up. We would like to take the saturation as a criteria for emergent spacetime (even though it is merely a necessary condition). At this point, this conjecture is mostly based on physical intuition. Hopefully, more evidences can be found in future studies.     

Ideally, we would like to compute the size or complexity of an operator in a CFT. There are quite a few available approaches and some of them are applicable to our current scheme. For example, it is straightforward to compute the complexity of a bulk operator using the techniques in \cite{Chagnet:2021uvi}. Roughly speaking, it is the cost to move a bulk operator radially to a given point from the boundary. Unfortunately, we find that the complexity grows linearly with the proper distance (instead of exponentially). We suspect the mismatch is due to the fact that the bulk operator (in the standard HKLL form, or for computation purpose the form in \cite{Goto:2017olq}) is essentially free (not seeing even the gravity). Another option is to compute the Krylov complexity (see \eg \cite{Parker:2018yvk, Kar:2021nbm, Caputa:2021ori}).

Chaos phenomenon has been studied extensively and there are many other measures of chaotic behavior. Hopefully, a better understanding of their imprints on the entanglement renormalization can help us patch together a more complete picture of spacetime emergence.     

\section*{Acknowledgments}
Xing Huang acknowledges the support of the NSFC Grants No. 11947301 and No. 12047502.

\bibliography{bio}

\end{document}